\documentclass{jpsj-suppl}
\usepackage{txfonts} 

\title{Effects of Nuclear Medium on the Sum Rules in Electron and Neutrino Scattering}
\author{F. \textsc{Zaidi}$^{1}$, H. \textsc{Haider}$^{1}$, M. Sajjad \textsc{Athar}$^{1}$, S. K. \textsc{Singh}$^{1}$ and I. \textsc{Ruiz Simo}$^{2}$}

\inst{$^{1}$Department of Physics, Aligarh Muslim University, Aligarh, 202002, India, $^2$Departamento de F\'{\i}sica At\'omica, Molecular y Nuclear,
and Instituto de F\'{\i}sica Te\'orica y Computacional Carlos I,
Universidad de Granada, Granada 18071, Spain }

\email{zaidi.physics@gmail.com}

\recdate{January 19, 2016}

\abst{In this work, we study the influence of nuclear medium effects on various parton model sum rules in nuclei 
and compare the results with the free nucleon case. We have used relativistic nucleon spectral function to take into account Fermi motion, binding and
nucleon correlations. The pion and rho meson cloud contributions have been incorporated in a microscopic model. The effect of shadowing has also been 
considered.}
\kword{parton-model sum rules, deep inelastic scattering, nuclear medium effect}
\begin{document}
\maketitle
\section{Introduction}
In the deep inelastic scattering, the target nucleon is considered to be a collection of quarks 
and gluons(described as partons). 
The deep inelastic cross sections for charged lepton-nucleon scattering or (anti)neutrino-nucleon scattering are 
 described in terms of nucleon structure functions which depend upon the
momentum distribution of these quarks and gluons. Using the appropriate relations between 
these structure functions one can obtain certain relations. These relations are better known as parton-model sum rules. Some of these
sum rules are Gross-Llewellyn Smith sum rule(GLS)~\cite{Gross:1969jf}, Adler sum rule(ASR)~\cite{Adler:1965ty},
 Gottfried sum rule(GSR)~\cite{Gottfried:1967kk}.
 
 GLS~\cite{Gross:1969jf} is defined for an isoscalar nucleon target 'N' and a symmetric sea as
\begin{equation}\label{eq1}
 S_{GLS}=\int_0^1 F_3^{\nu_lN} (x) dx=3.
\end{equation}

 ASR~\cite{Adler:1965ty} predicts the difference between the quark densities 
of the neutron and the proton, and is given by
\begin{equation}
S_{ASR}=\int_0^1 \frac{dx}{x} [F_2^{\nu_l n} (x) - F_2^{\nu_l p} (x) ]=2.
\label{eq2}
\end{equation}

GSR~\cite{Gottfried:1967kk} also known as valence isospin sum rule is given by
\begin{equation}
S_{GSR}=\int_0^1 \frac{dx}{x} [F_2^{ep} (x) - F_2^{en} (x) ]=\frac{1}{3}~+~\frac{2}{3}\int_0^1 dx~ (\bar u - \bar d)
\label{eq3}
\end{equation}
  where $F_2^{eN}(x)$ is the electromagnetic nucleon structure function and $F_i^{\nu_l(\bar\nu_l)N}(x);~l=e,\mu$, i=2,3 is the weak nucleon structure functions.
 
With the development of high precision neutrino/antineutrino experiments as well as high luminosity electron beam experiments 
 it is possible to verify these sum rules. These experiments are being done with moderate
and heavier nuclear targets. For a nucleus, these sum rules are expressed in terms of nuclear structure
functions like $F_{1A}^{EM}(x,Q^2)$ and $F_{2A}^{EM}(x,Q^2)$ for electromagnetic processes
and $F_{1A}^{Weak}(x,Q^2)$, $F_{2A}^{Weak}(x,Q^2)$ and $F_{3A}^{Weak}(x,Q^2)$ for weak interaction induced processes, which get modified because 
of the nucleons bound inside the nucleus. In the present work, we have taken into account 
nuclear medium effects like Fermi motion, binding energy, nucleon correlations, etc., using a relativistic nucleon 
spectral function in an interacting Fermi sea and local density approximation is then applied to obtain the results for finite nuclei.
 Furthermore, mesonic contributions and shadowing effects have also been taken into account.
 The results are compared with the free nucleon case as well as with some of the available experimental data. 
 
 The details of the present formalism
 are given in Ref.\cite{Haider:2015vea}. We are presenting the formalism in brief.
\section{Formalism}
For the charged lepton induced deep inelastic scattering process 
($l(k) + N(p) \rightarrow l(k^\prime) + X(p^\prime);$ $l=~e^-,~\mu^-$), 
the differential scattering cross section is given by
\begin{equation}\label{eA}
\frac{d^2 \sigma^N}{d\Omega_l dE_l^{\prime}} =~\frac{\alpha^2}{q^4} \; \frac{|\bf k'|}{|\bf k|} \;L_{\mu \nu} \; W_N^{\mu \nu},
\end{equation}
where the hadronic tensor $W_N^{\mu \nu}$ is defined in terms of nucleon structure functions $W_i^N$(i=1,2) as

\begin{equation}\label{nuclearht}
W_N^{\mu \nu} = 
\left( \frac{q^{\mu} q^{\nu}}{q^2} - g^{\mu \nu} \right) \;
W_1^N + \left( p_N^{\mu} - \frac{p_N . q}{q^2} \; q^{\mu} \right)
\left( p_N^{\nu} - \frac{p_N . q}{q^2} \; q^{\nu} \right)
\frac{W_2^N}{M^2}
\end{equation}
with $M$ as the mass of nucleon.

 For the lepton scattering taking place with a nucleon moving inside the nucleus, the expression of the  cross section is modified as
\begin{equation}\label{eA}
\frac{d^2 \sigma^A}{d\Omega_l dE_l^{\prime}} =~\frac{\alpha^2}{q^4} \; \frac{|\bf k'|}{|\bf k|} \;L_{\mu \nu} \; W_A^{\mu \nu},
\end{equation}
where $W_A^{\mu \nu}$ is the nuclear hadronic tensor defined in terms of nuclear hadronic structure functions $W_i^A$(i=1,2) as

\begin{equation}\label{nuclearht}
W_A^{\mu \nu} = 
\left( \frac{q^{\mu} q^{\nu}}{q^2} - g^{\mu \nu} \right) \;
W_1^A + \left( p_A^{\mu} - \frac{p_A . q}{q^2} \; q^{\mu} \right)
\left( p_A^{\nu} - \frac{p_A . q}{q^2} \; q^{\nu} \right)
\frac{W_2^A}{M_A^2}
\end{equation}
with $M_A$ as the mass of nucleus.

To get $d\sigma$ for $(l, l')$ scattering on the nucleus, we are required to evaluate
imaginary part of lepton self energy $\Sigma (k)$ which is written using Feynman rules as~\cite{Haider:2015vea}
\begin{equation}
\Sigma (k) = i e^2 \; \int \frac{d^4 q}{(2 \pi)^4} \;
\frac{1}{q^4} \;
\frac{1}{2m} \;
L_{\mu \nu} \; \frac{1}{k'^2 - m^2 + i \epsilon} \; \Pi^{\mu \nu} (q),
\end{equation}
where $\Pi^{\mu \nu} (q)$ the photon self energy and $L_{\mu \nu}$ is the leptonic tensor 
 $L_{\mu \nu} = 2 (k_{\mu} k'_{\nu} +  k'_{\mu} k_{\nu} - k\cdot k^\prime g_{\mu \nu})$.
Now we shall use the imaginary part of the lepton self energy  i.e. $Im \Sigma (k)$, to obtain the results for the cross section and for this we apply 
Cutkosky rules
\begin{equation}\label{cut}
\begin{array}{lll}
\Sigma (k) & \rightarrow & 2 i \; I m \Sigma (k),~~~~~~~~~~~~~~~~~~D (k') \rightarrow  2 i \theta (k'^0) \; I m D (k') \\
\Pi^{\mu \nu} (q) & \rightarrow & 2 i \theta (q^0) \; I m \Pi^{\mu \nu} (q), ~~~~~~~G (p) \rightarrow  2 i \theta (p^0) \; I m G (p) 
\end{array}
\end{equation}
which leads to
\begin{eqnarray}\label{self-lepton}
Im \Sigma (k) &=& e^2 \int \frac{d^3 q}{(2 \pi)^3} \;
\frac{1}{2E_l}\theta(q^0) \; Im(\Pi^{\mu\nu})  \frac{1}{q^4} \frac{1}{2m} \; L_{\mu \nu}\;
\end{eqnarray}

 Notice from Eq.~\ref{self-lepton}, $\Sigma (k)$ contains photon self energy $\Pi^{\mu\nu}$, which is 
written in terms of nucleon  propagator $G_l$ and meson  propagator $D_j$ and using Feynman rules this is given by
\begin{eqnarray}\label{photonse}
\Pi^{\mu \nu} (q)&=& e^2 \int \frac{d^4 p}{(2 \pi)^4} G (p) 
\sum_X \; \sum_{s_p, s_l} {\prod}_{\substack{i = 1}}^{^N} \int \frac{d^4 p'_i}{(2 \pi)^4} \; \prod_{_l} G_l (p'_l)\; \prod_{_j} \; D_j (p'_j)\nonumber \\  
&&  <X | J^{\mu} | H >  <X | J^{\nu} | H >^* (2 \pi)^4  \; \delta^4 (q + p - \sum^N_{i = 1} p'_i),\;\;\;
\end{eqnarray}
where $s_p$ is the spin of the nucleon, $s_i$ is the spin of the fermions in $X$, $<X | J^{\mu} | H >$ is the hadronic current for the initial state nucleon 
to the final state hadrons, index $l$ runs for fermions and index $j$ runs for bosons in the final hadron state $X$. 
 
 The relativistic nucleon propagator G(p) in a nuclear medium is obtained as~\cite{FernandezdeCordoba:1991wf,Marco:1995vb}:
\begin{eqnarray}\label{Gp}
G (p) =&& \frac{M}{E({\bf p})} 
\sum_r u_r ({\bf p}) \bar{u}_r({\bf p})
\left[\int^{\mu}_{- \infty} d \, \omega 
\frac{S_h (\omega, {\bf{p}})}{p_0 - \omega - i \eta}
+ \int^{\infty}_{\mu} d \, \omega 
\frac{S_p (\omega, {\bf{p}})}{p_0 - \omega + i \eta}\right]\,,
\end{eqnarray}
where $S_h (\omega, {\bf{p}})$ and $S_p (\omega, {\bf{p}})$ being the hole
and particle spectral functions respectively, which are given in Ref.\cite{FernandezdeCordoba:1991wf}.
 
 The cross section is then obtained as: 
\begin{equation}\label{dsigma_3}
\frac {d\sigma^A}{d\Omega_l dE_l'}=-\frac{\alpha}{q^4}\frac{|\bf{k^\prime}|}{|\bf {k}|}\frac{1}{(2\pi)^2} L_{\mu\nu} \int  Im \Pi^{\mu\nu}d^{3}r
\end{equation}

After performing some algebra, the expression of the nuclear hadronic tensor for an isospin symmetric nucleus in terms of 
 nucleonic hadronic tensor and spectral function, is obtained as~\cite{Haider:2015vea}
\begin{equation}	\label{conv_WA}
W^{\alpha \beta}_{A} = 4 \int \, d^3 r \, \int \frac{d^3 p}{(2 \pi)^3} \, 
\frac{M}{E ({\bf p})} \, \int^{\mu}_{- \infty} d p_0 S_h (p_0, {\bf p}, \rho(r))
W^{\alpha \beta}_{N} (p, q), \,.
\end{equation}

Accordingly the dimensionless nuclear 
structure functions $F_{i=1,2}^A(x,Q^2)$, are defined in terms of $W_{i=1,2}^A(\nu,Q^2)$ as
\begin{eqnarray}\label{relation1}
F_1^A(x,Q^2)&=&M_A~W_{1}^{A}(\nu,Q^2)~\nonumber \\
F_2^A(x,Q^2)&=&\nu_A~W_{2}^{A}(\nu,Q^2)~  {\rm where}\nonumber \\
\nu_A&=&\frac{p_{_A}\cdot q}{M_{_A}}=\frac{p_{0_A} q_{0}}{M_{_A}}=q_{0},~~ p_{_A}^\mu=(M_{_A},\vec 0) {\rm ~and~} M_{_A} {\rm ~is~ the~ mass~ of~ a~ nucleus.}
\end{eqnarray}

For weak interaction, we follow the same procedure, formalism for which is given in accompanying paper by Haider et al.~\cite{Haider} in this proceeding.  
 For a nonisoscalar nuclear target the expression for the dimensionless structure functions $F_{_{1~A}}(x_A, Q^2)$ and $F_{_{2~A}}(x_A, Q^2)$ are obtained as
\begin{eqnarray}	\label{conv_WA1}
F_{_{1~A}}^{EM/Weak}(x_A, Q^2) &=& 2\sum_{\tau=p,n} AM \int \, d^3 r \, \int \frac{d^3 p}{(2 \pi)^3} \, 
\frac{M}{E ({\bf p})} \, \int^{\mu}_{- \infty} d p_0 S_h^\tau (p_0, {\bf p}, \rho^\tau(r))~\left[\frac{F_{1}^{EM/Weak,\tau}(x_N, Q^2)}{M}\right. \nonumber\\
&& \left. + \frac{1}{M^2}{p_x}^2 \frac{F_{2}^{EM/Weak,\tau}(x_N, Q^2)}{\nu}\right].~~~
\end{eqnarray}

\begin{eqnarray} \label{had_ten151}
F_{_{2~A}}^{EM/Weak}(x_A, Q^2)  &=&  2\sum_{\tau=p,n} \int \, d^3 r \, \int \frac{d^3 p}{(2 \pi)^3} \, 
\frac{M}{E ({\bf p})} \, \int^{\mu}_{- \infty} d p_0 S_h^\tau (p_0, {\bf p}, \rho^\tau(r)) \times\left[\frac{Q^2}{q_z^2}\left( \frac{|{\bf p}|^2~-~p_{z}^2}{2M^2}\right)\right. \nonumber \\
&& \left. +  \frac{(p_0~-~p_z~\gamma)^2}{M^2} \left(\frac{p_z~Q^2}{(p_0~-~p_z~\gamma) q_0 q_z}~+~1\right)^2\right]~\frac{M}{p_0~-~p_z~\gamma} ~F_2^{EM/Weak,\tau}(x,Q^2)
,~~\;\;\;~~       
\end{eqnarray}
where $\gamma=\frac{q_z}{q_0}$.

\begin{eqnarray}
F_{_{3~A}}^{Weak}(x_A,Q^2)&=&2\int d^3r \int \frac{d^3p}{(2\pi)^3} \frac{M}{E({\vec p})}\left[\int_{-\infty}^{\mu} dp^0
S_h^p(p^0,\mathbf{p},k_{F,p}) F_3^p(x_N,Q^2) \right. \nonumber \\
&& \left. + \int_{-\infty}^{\mu} dp^0
S_h^n(p^0,\mathbf{p},k_{F,n}) F_3^n(x_N,Q^2) \right] \times \left( \frac{p^0\gamma-p_z}{(p^0-p_z\gamma)\gamma}\right) .
\end{eqnarray}

The nucleon structure functions $F_i^N(x,Q^2)$ (i=1-3), are expressed in terms of parton distribution functions(PDFs), for which we have taken the parameterization of
CTEQ6.6~\cite{cteq}. The evaluations are performed at the Leading-Order(LO) as well as Next-to-Leading-Order(NLO). In the evaluation of 
sum rules(Eqs.\ref{eq1}-\ref{eq3}), the mesonic contribution cancels out if we follow the parameterization of Gluck et al.~\cite{Gluck} for pion PDFs.
\section{Results}
\begin{figure}[tbh]
\includegraphics[height=5 cm, width=12 cm]{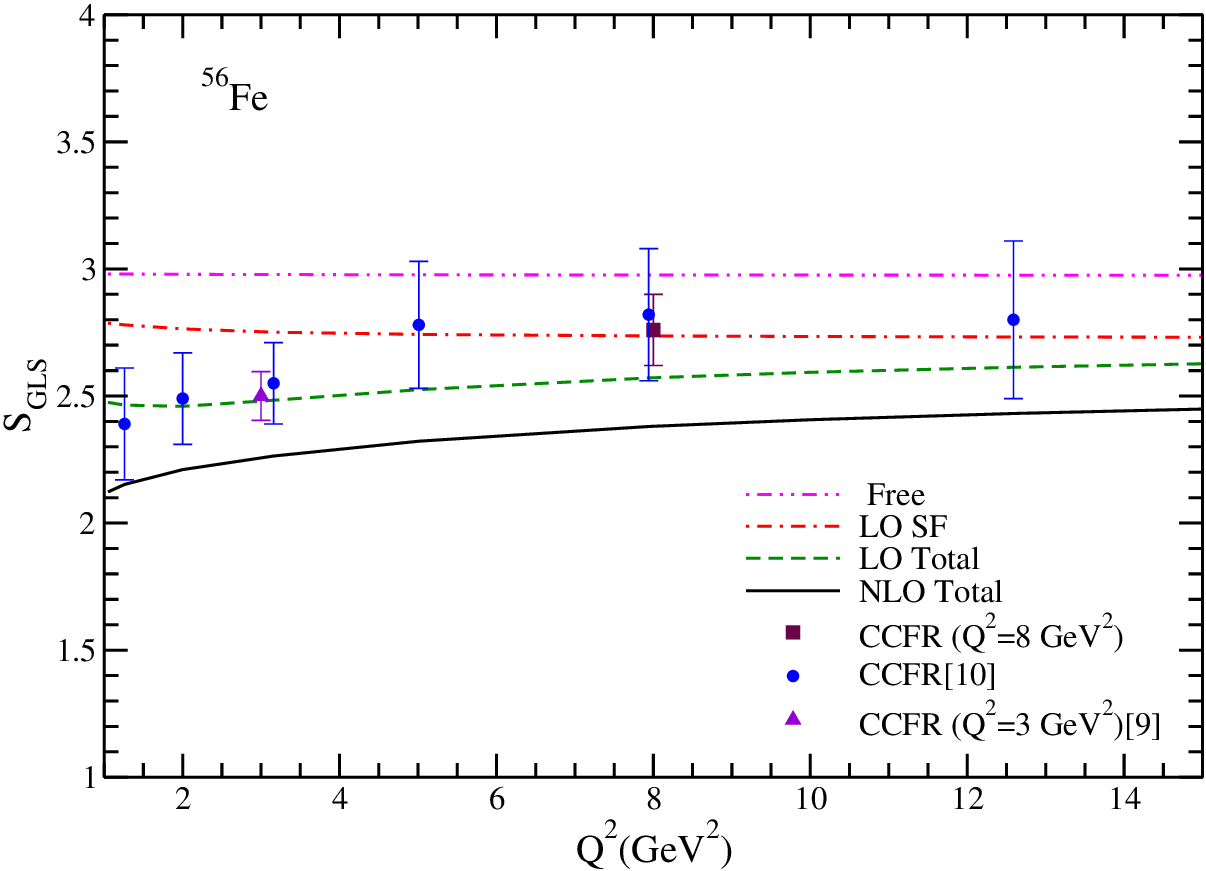}
\caption{Results for GLS sum rule in $^{56}Fe$ at both LO and NLO and the
results are also compared with CCFR experimental data~\cite{Leung:1992yx, Kim:1998kia}.}
\label{f2}
\end{figure}

\begin{figure}
\includegraphics[height=5 cm, width=8 cm]{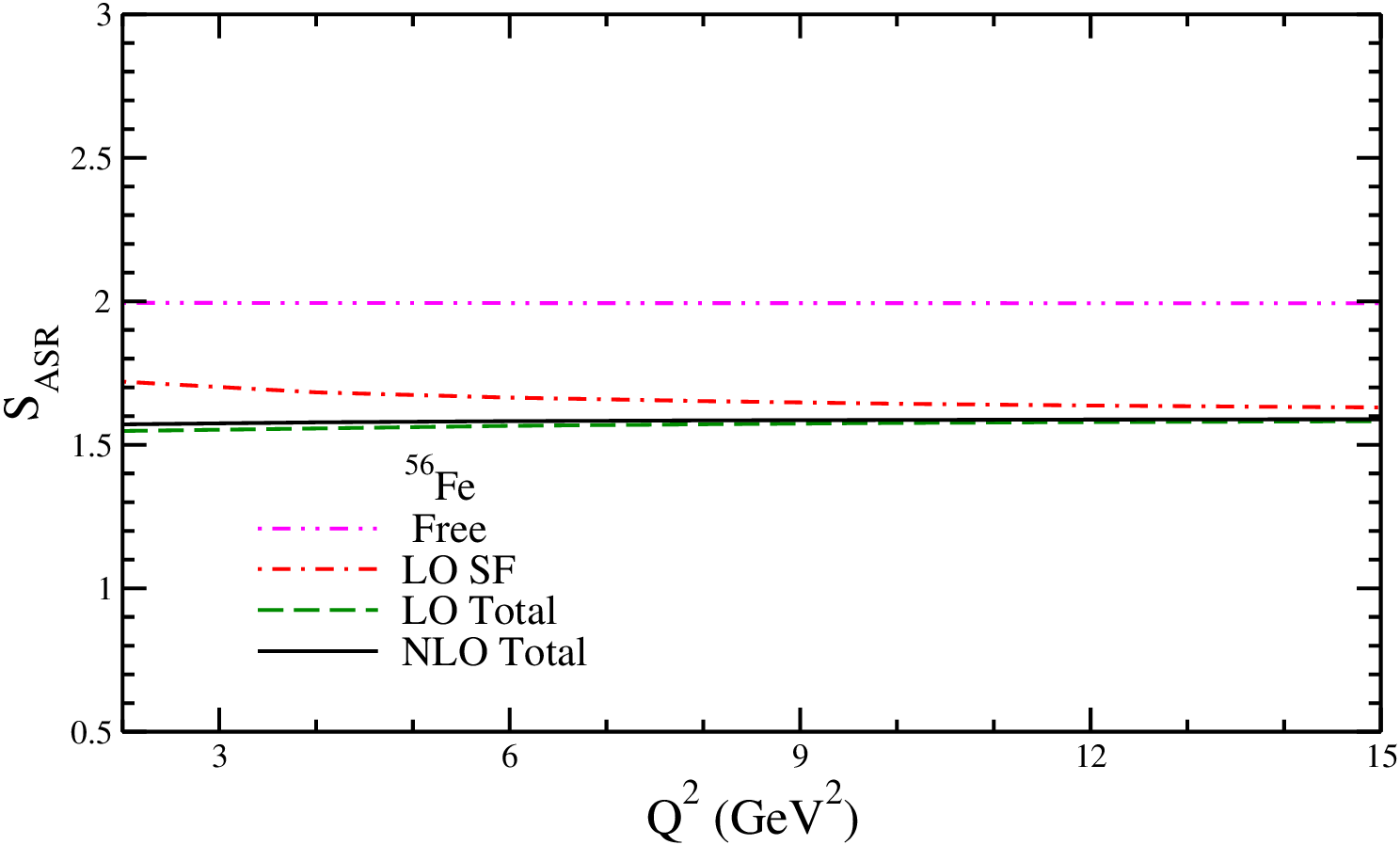}
\includegraphics[height=5 cm, width=8 cm]{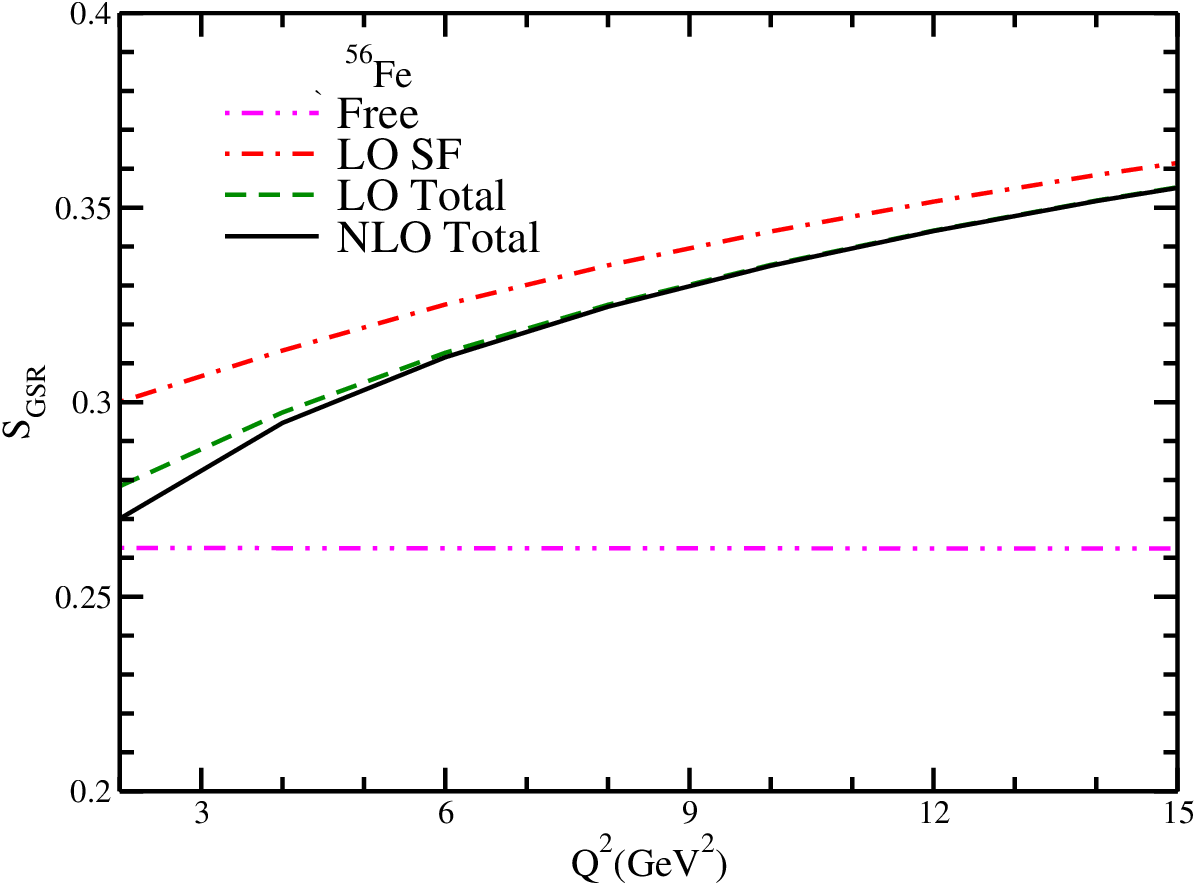}
\caption{{\bf Left panel}:Results for Adler sum rule in $^{56}Fe$ at both LO and NLO and the results are also
compared with the results of free nucleon. {\bf Right panel}:Results for Gottfried sum rule in $^{56}Fe$ at both LO and NLO.}
\label{f3}
\end{figure}

In Fig.\ref{f2}, we have presented the results for the GLS sum rule in the free nucleon evaluated at LO as well as in $^{56}Fe$ nucleus, using spectral
 function of the nucleon. We find that when calculations are done using spectral function, the value of GLS integral 
 decreases from the free nucleon case which is around $7-8\%$ at all values of $Q^2$. When shadowing effects are included following Ref.~\cite{Petti2} (note that 
 there is no mesonic contribution to $F_3^N(x,Q^2)$), $S_{GLS}$ further reduces by $\sim 10\%$ at low $Q^2$ and about
 $3-4\%$ at $Q^2=12-15~ GeV^2$. This is the result of our full calculation at LO. When  we evaluate the results at NLO, $S_{GLS}$ further decreases by 
 $\sim 10\%$ at low $Q^2$ and $6\%$ at $Q^2=10-15 ~GeV^2$.

Similarly in Fig. \ref{f3},  the results are presented for Adler and Gottfried sum rules. We find that $S_{ASR}$
decreases from free nucleon case when spectral function is used, and the decrease is $\sim 14-18\%$ in $2 ~GeV^2 ~<~Q^2~< 15~ GeV^2$.
 When shadowing  and mesonic effects are included there is further reduction of $\sim 10\%$ at low $Q^2$ and $\sim 3-4\%$ at high $Q^2$.
 The evaluation at NLO results in a very small change in the Adler sum rule.
 
 In the right panel of this figure, we show the results for GSR evaluated with four quark flavors(u,d,s,c).
 We find that the inclusion of spectral function results
 in an increase of $S_{GSR}$ which is $\sim 14-15\%$ at $Q^2=2-3~ GeV^2$ which becomes $\sim 30-35\%$ 
 at $Q^2=10-15 ~GeV^2$ from free nucleon case.
 However, when shadowing and mesonic effects are taken into account, the net increase in $S_{GSR}$ is 7-8$\%$ at low $Q^2$( for $Q^2=2-3~ GeV^2$) which becomes 30-32$\%$ at higher $Q^2$ (for $Q^2=10-15 ~GeV^2$).
 There is 2-3$\%$ reduction at low $Q^2$ when the calculations are performed at NLO which becomes negligible at high $Q^2$.

 We conclude that there is significant dependence of nuclear medium effects in the sum rules studied in this work. 
 Morever, we find that nuclear medium effects lead to $Q^2$ dependence in these sum rules. This study may be useful in the future analysis of experiments looking for the 
validity of sum rules.

\end{document}